\documentclass{cs20proc}

\usepackage{kantlipsum,natbib}
\usepackage{hyperref}
\hypersetup{
    colorlinks=true,
    linkcolor=blue,
    filecolor=magenta,      
    urlcolor=blue,
    pdftitle={Overleaf Example},
    pdfpagemode=FullScreen,
    citecolor=blue
    }
\editors{A. S. Brun, J. Bouvier, P. Petit}
\publisher{Zenodo}
\conference{The 21th Cambridge Workshop on Cool Stars, Stellar Systems, and the Sun}
\conferencedate{2022}

\title{Space Weather Research using Spectropolarimetric Radio Imaging Combined With Aditya-L1 and PUNCH Missions}
\author{Devojyoti Kansabanik,$^{1}$
        Surajit Mondal$^{2}$,
        Divya Oberoi $^1$, Puja Majee $^1$}

\affiliation{$^{1}$ National Centre for Radio Astrophysics, Tata Institute of Fundamental Research, S. P. Pune University Campus, Pune 411007, India \\
$^{2}$ Center for Solar-Terrestrial Research, New Jersey Institute of Technology, 323 M L King Jr Boulevard, Newark, NJ 07102-1982, USA}

\shorttitle{Space-weather using Radio Imaging}
\shortauthors{Kansabanik et al.}

\abs{Low-frequency radio observations have been expected to serve as a powerful tool for Space Weather (SW) observations for decades. Radio observations are sensitive to a wide range of SW-related observations ranging from emissions from coronal mass ejections (CMEs) to the solar wind. Ground-based radio observatories allow one gathering of high-sensitivity data at high time and spectral resolution for an extended period, which remains a challenge for most space-based observatories. While radio techniques like Interplanetary Scintillation (IPS) are well established, radio imaging studies have remained technically challenging. This is now changing with the confluence of data from instruments, like the Murchison Widefield Array (MWA), and robust unsupervised analysis pipelines. This pipeline delivers full Stokes radio images with unprecedented fidelity and dynamic range. This will serve as a powerful tool for coronal and heliospheric studies. We present the recent developments and achievements to measure the magnetic fields of the CME plasma and shock front at coronal heights and also share the current status of the objective to measure the heliospheric Faraday rotation towards numerous background linearly polarised radio sources with the Sun in the field of view. We envision that in the coming years, the availability of new-generation radio instruments combined with the Aditya-L1 and PUNCH mission will mark the start of a new era in Space Weather modeling and prediction.}

\begin{document}

\maketitle

\section{Introduction}
The space weather around the Earth is determined by the Sun. The most important phenomenon determining the space weather is coronal mass ejection (CME). CMEs are large-scale eruptions of magnetized plasma from the Sun into the heliosphere. It is well-established that the magnetic fields play important roles in their propagation and determining their geo-effectiveness. While propagating CMEs interact with other heliospheric components like solar wind, co-rotating interaction regions, and stream interaction regions and change their propagation direction and magnetic field topology \citep{Manchester2017}. These deformations complicate the prediction of CME arrival times or the $B_\mathrm{z}$ component of the magnetic field at 1 AU. Hence, tracking and measuring the magnetic fields of a CME as it propagates from the corona into the heliosphere, is essential for improving space-weather forecasting.

There are several state-of-the-art CME models \citep{Isavnin_2016,Möstl2018} developed over the last few years to incorporate these deformations of CMEs into account. These models have multiple independent parameters. One needs to constrain these model parameters of a CME during its propagation from lower coronal heights to the inner heliosphere so that accurate space weather predictions could be made. White-light coronagraphs and heliospheric imagers provide routine observations of the CME structures, which allow us to measure the geometrical and dynamical properties of the CMEs and in certain cases, magnetic fields measurements at the CME shock fronts \citep{Gopalswamy2011}, but can not provide any direct measurement of the magnetic fields of the CME plasma. To date, direct measurements of the magnetic field and other plasma parameters of the CME plasma can be obtained only by using {\it in-situ} observations from different vantage points in space, using multiple spacecrafts. But, for accurate space-weather prediction, one needs remote measurements of the CME-entrained magnetic fields at the coronal and heliospheric heights. 

Over the past several years, multiple new instruments and new mission concepts have materialized at all wavelengths (X-rays to radio) to remotely measure different properties of the CMEs, particularly their vector magnetic fields.  Among all wavelengths, radio observations are particularly well-suited for the remote measurements of the CME magnetic fields. In this article, we describe opportunities presented by the recent developments in radio observations for space-weather research using a new-technology instrument, the {\it Murchison Widefield Array} \citep[MWA,][]{lonsdale2009,Tingay2013,Wayth2018} and the synergies between the MWA observations with the upcoming first Indian solar mission; Aditya-L1 \citep{Tripathi2017} and the future PUNCH mission \citep{punch2022}.

We organize this paper as follows. We describe the space-weather observable at radio wavelengths in Section \ref{sec:radio_observables}. In Section \ref{sec:challenges_radio} we discuss the challenges in radio observations, followed by a discussion of the recent achievements in overcoming these challenges in Section \ref{sec:radio_results}. We then discuss the importance of joint observations with the Aditya-L1 in Section \ref{label:aditya} followed by the synergies with the PUNCH mission in Section \ref{label:punch}. We conclude the work in Section \ref{sec:conclusion}.

\section{Space-weather Observable at Radio Wavelengths}\label{sec:radio_observables}
All kinds of eruptive phenomena in the solar corona, flares \citep{Cargill2000} to CMEs \citep{KAHLER20032587}, are efficient particle accelerators, either due to magnetic re-connection or shocks. These energetic particles, particularly the electrons, produce different kinds of radio emissions through different emission mechanisms. Among them, coherent plasma emission and in-coherent gyrosynchrotron/ thermal emission are two important space-weather observables. 

\begin{figure}
    \centering
    \includegraphics[trim={0cm 0cm 0cm 0cm},clip,scale=0.3]{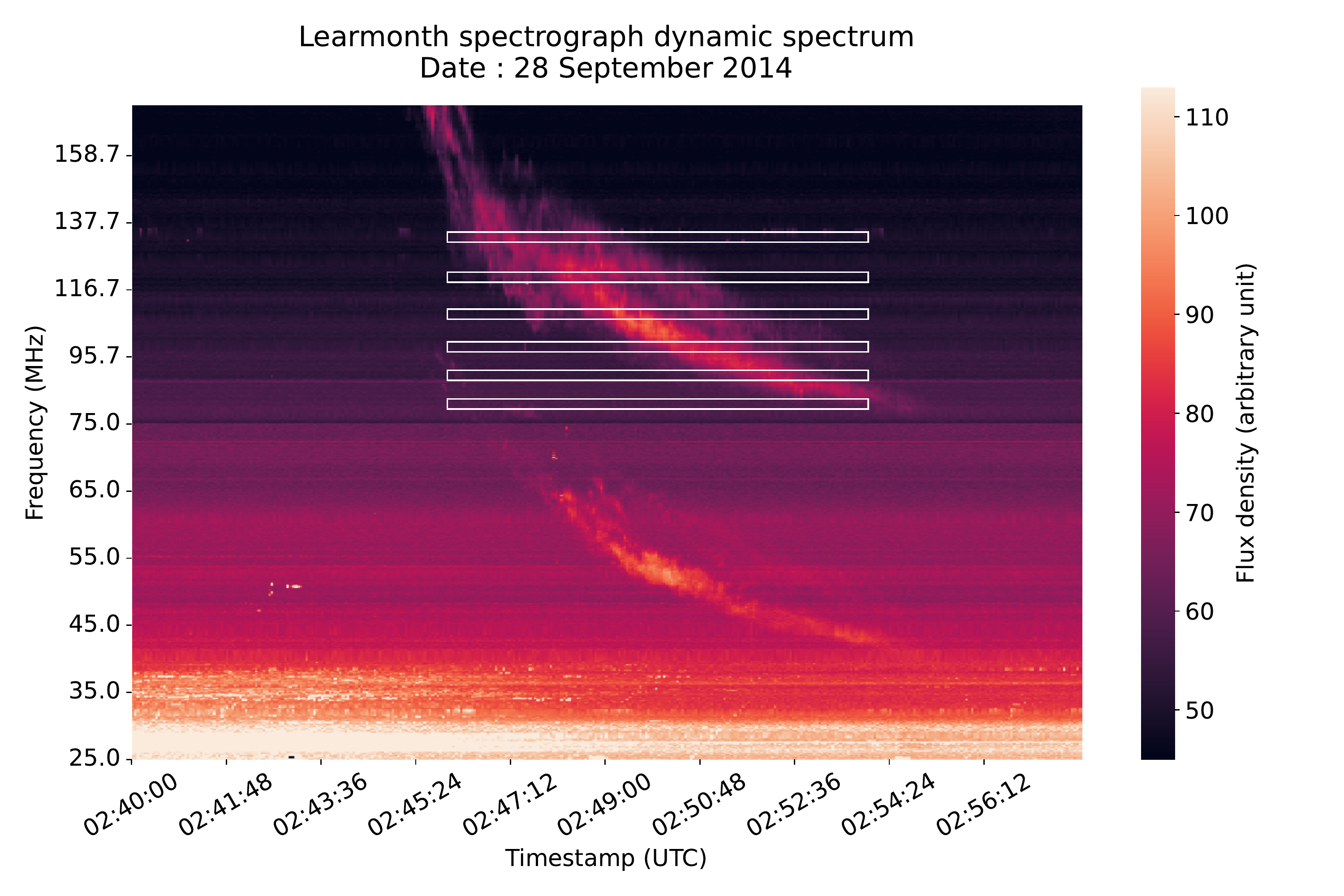}
    \vspace{-0.5cm}
    \caption{An example of a type-II solar radio burst was observed using Learmonth Solar Spectrograph on 2014-September-28. The rectangular boxes represent the part of the dynamic spectrum with simultaneous MWA observations.}
    \label{fig:type-II}
    \vspace{-0.5cm}
\end{figure}

Coherent plasma emissions are classified into different types based on their appearance in the dynamic spectrum \citep{McLeanBook}. Among these different types of radio bursts, type-II radio bursts are directly linked to coronal shocks either generated by CMEs or other eruptive events \citep{Suli2020}. Type-II radio bursts appear as slowly drifting features in the dynamic spectrum (Figure \ref{fig:type-II}) and their brightness temperature ($T_B$) can vary between $10^6-10^{12}$ K. Non-imaging polarization observations \citep[e.g.,][etc.]{Ramesh_2022,Ramesh_2023} and band-splitting \citep[e.g.,][etc.]{Vrsnak2001,Cho2007,MAHROUS201875} of the type-II radio bursts have been used to measure the average magnetic field strength at the shock front of the CMEs. Although successful, these non-imaging observations sometimes have ambiguity in identifying whether the band-split type-II emission is coming from the same shock or different shocks. Spectro-polarimetric imaging observations are important to localize the type-II emission at the shock front, which can remove such ambiguities. 

Type-II bursts can occur at any coronal height and even in interplanetary space \citep[e.g.,][etc.]{Krupar2015,Jebaraj2021}. Because the emission frequency is proportional to the square root of the local electron density, the emission frequency of the type-II radio bursts at higher coronal heights is below the ionospheric cutoff frequency ($\sim10$ MHz). Hence, the imaging observations are only possible using the ground-based instruments below $\sim2\ R_\odot$. In the future, space-based instruments like the Sun Radio Interferometer Space Experiment \citep[SunRISE,][]{Romero2020,Lazio2021} can also perform imaging localization of the type-II radio bursts at higher coronal heights and interplanetary space.

Type-II radio bursts can provide magnetic field measurements at the shock front, but can not be used to remotely measure the CME-entrained magnetic fields. There are two possible methods to measure the CME-entrained magnetic fields using radio observations. These two methods are gyrosynchrotron (GS) emission produced by the mildly-relativistic electrons \citep[e.g.,][etc.]{bastian2001,Mondal2020a} and induced circular polarization from thermal free-free emission \citep[e.g.,][etc.]{Gopalswamy1993,Ramesh2021} in the presence of CME magnetic fields. Using the ground-based radio observations, these methods can be used to measure the CME-entrained magnetic fields up to $\sim10\ R_\odot$ \citep{Kansabanik2023_CME1}.

Beyond $\sim10\ R_\odot$, both the GS or thermal radio emissions from the CME plasma becomes too faint to detect using the current generation instruments and also the optimal observing frequency becomes smaller than the ionospheric cutoff. Although the observation of direct radio emission from the CME plasma is not possible at higher coronal heights and the inner heliosphere, two in-direct radio observables can be used to measure the CME plasma properties at these heights. These two methods are interplanetary scintillation (IPS) \citep[see,][for a review]{Briggs1966,Coles1978} and Faraday rotation (FR) measurements \citep[see,][for a review]{Kooi2022} of background galactic/extra-galactic radio sources. IPS observations have routinely been used to measure the velocity and density fluctuations in the heliosphere. On the other hand, FR measurements of background linearly polarized radio sources can be used to measure the line-of-sight (LoS) integrated magnetic field of the CME and solar wind. Both of these methods can help constrain state-of-the-art CME models like FRiED \citep{Isavnin_2016} or 3DCORE \citep{Möstl2018} and can be used to estimate vector magnetic fields of the CMEs at the higher coronal heights and inner heliosphere. 

\section{Challenges in Observation and Recent Developments}\label{sec:challenges_radio}
Despite the existence of several methods to remotely estimate the CME magnetic fields using ground-based radio observations, their usuage remains limited due to observational challenges both for the direct and indirect methods. Both the type-II radio bursts and GS/thermal emission from CME plasma show spectro-temporal and spatial variations at different scales \citep{Kansabanik_principle_AIRCARS}. More importantly, the $T_B$ and circular polarization (Stokes V) fraction also vary by several orders of magnitude \citep{Kansabanik_principle_AIRCARS}. $T_B$ of type-II emission can vary between $10^6-10^{12}$ K, the $T_B$ of GS/thermal emissions are several orders of magnitude smaller, $\sim10^3-10^4$ K. Very often it has been found that the faint GS/thermal sources are present simultaneously with the very bright coherent emissions. Hence to measure the spatially resolved magnetic fields using these methods, one needs to image them with high-dynamic range and high-fidelity at a high spectro-temporal cadence. 

The major limitations in high-dynamic-range and high-fidelity spectro-polarimetric imaging of the Sun come from the instrument and calibration. These problems have been resolved using one of the new-technology radio interferometric arrays, the MWA. The MWA is a radio interferometer array operating at 80-300 MHz and comprised of 128 antenna tiles (currently 144 antenna elements) distributed over a region of 5 km diameter. The dense array configuration of the MWA makes it well-suited for high-dynamic-range spectroscopic snapshot imaging. Over the past several years, calibration and imaging pipelines \citep{Mondal2019, Kansabanik2022,Kansabanik2022_paircarsI} have been developed for the MWA solar observations. ``Polarimetry using Automated Imaging Routine for the Compact Arrays for the Radio Sun" \citep[P-AIRCARS,][]{Kansabanik_paircarsII} is the current state-of-the-art full Stokes calibration and imaging pipeline. It is capable of routinely producing high-dynamic-range ($\sim10^3-10^5$) images with high-fidelity polarimetric calibration (residual instrumental leakage $\leq1\%$). P-AIRCARS is already leading to several successes, those related to space-weather research are discussed in the following Section \ref{sec:radio_results}.

\section{Recent Achievements at Radio Wavelengths}\label{sec:radio_results}
High-dynamic-range and high-fidelity spectro-polarimetric imaging provided by P-AIRCARS now allows us to explore the previously inaccessible phase-space. Some glimpses of new possible explorations are discussed below.

\subsection{Spectropolarimetric Imaging of Type-II Solar Radio Bursts}
Type-II solar radio bursts are coherent plasma emissions caused by magnetohydrodynamics (MHD) shocks driven by solar eruptive events such as Coronal Mass Ejections (CMEs), flares, and jets \citep{McLeanBook,Alissandrakis2021}. Despite  observations spanning several decades, type-II radio bursts are still not understood completely. Although there are several open questions about type-II radio bursts, several studies have been used to measure the magnetic field strength and turbulence at the shock front \citep[e.g.,][etc.]{McLeanBook,Cho2007,Ramesh_2022,Ramesh_2023}. Most of these studies used non-imaging observations and made several assumptions while estimating the shock properties from the observations. Spectro-polarimetric snapshot imaging observations of type-II radio bursts allowed us to understand some of these mysteries and examine these assumptions in detail.

\begin{figure}
    \centering
    \includegraphics[trim={0cm 0cm 0cm 0cm},clip,scale=0.28]{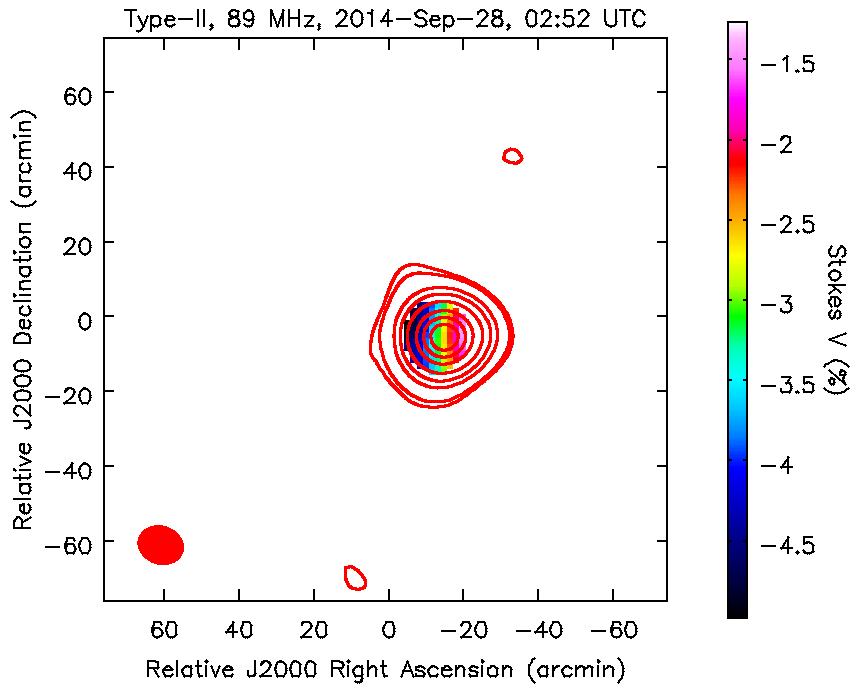}
    \caption{Stokes V emission from type-II solar radio burst observed using the MWA on 2014-September-28, 02:52 UTC. The color map shows the Stokes V percentage where Stokes V is detected with more than 5$\sigma$ significance. Red contours represent the Stokes I emission. Contours are at 0.5, 1, 5, 10, 20, 40, 60, and 80 \% of the peak Stokes I flux density. The blue dotted line represents the optical disc of the Sun. The red-filled ellipse at the bottom left is the point spread function of the instrument.}
    \vspace{-0.5cm}
    \label{fig:type-II_V}
\end{figure}

\cite{Bhunia2022} have studied a type-II radio burst observed with the MWA on 2014-September-28. The dynamic spectrum of this type-II radio bursts observed using the Learmonth Solar Spectrograph is shown in Figure \ref{fig:type-II}. The rectangular boxes show the overlapping MWA imaging observations with the MWA. Very often type-II band-splitting is associated with the shock upstream-downstream scenario \citep{Smerd1974} to predict the magnetic field strengths \citep{Cho2007} at the CME shock front. But the imaging observation by \cite{Bhunia2022} demonstrated that not only the source locations of the upper and lower bands do not match, these sources also move in different directions with different plane-of-the-sky speeds. This imaging observation suggests that band-splitting is caused by emission from multiple parts of the shock. 

One may suspect this conclusion by the fact that these could happen due to propagation effects. But the MWA observing frequency only covered the harmonic emission of the type-II, as evident from the Learmonth dynamic spectrum in Figure \ref{fig:type-II}. At harmonic frequency, the propagation effects are much less compared to fundamental and could not produce the observed spatial differences between the upper and lower band sources. Further investigations have been carried out using the full Stokes images provided by P-AIRCARS. An example fractional Stokes V image of a single time and frequency slice is shown in Figure \ref{fig:type-II_V}. The peak circular polarization fractional is small, $\sim5\%$. This small polarization fraction is consistent with previous studies, which have found that the fundamental emission generally has a higher polarization fraction and harmonic shows a lower polarization fraction \citep[e.g.,][]{Ramesh_2022}. In this case, one can not use the band-splitting to estimate the magnetic field of the CME shock front, but the polarization measurement can be used to directly measure the magnetic field. 

Following \cite{Melrose1980} and \cite{Zlotnik1981}, the degree of circular polarization for the harmonic plasma emission is given by 
\begin{equation}
    \mathrm{dcp}\approx \frac{11}{48}\frac{f_\mathrm{B}|cos\theta|}{f_\mathrm{p}}
    \label{eq:1}
\end{equation}
where, $f_\mathrm{B}$ (MHz) = 2.8$B$ is the electron gyro-frequency, $f_\mathrm{p}$ (MHz) is the plasma frequency and $\theta$ is the angle between magnetic field direction and LoS. The approximate expression of degree of circular polarization provided in Equation \ref{eq:1} is valid for circular polarization fraction of less than 10\%, which is valid in this case. $\theta$ can be approximated to the heliographic longitude \citep{Dulk1980,Ramesh_2022} of the centroid of the type-II burst. In this case, $\theta$ is $\sim60^\circ$. For harmonic emission, $f_\mathrm{p}$ = 44.5 MHz. Using these values and considering the spatially averaged degree of circular polarization of 3\% in Equation \ref{eq:1}, we have estimated the magnetic field at the shock front is $\sim2.2$ G. This demonstrates the importance of spectro-polarimetric imaging observation to estimate the magnetic field at the CME shock front using type-II solar bursts. With snapshot imaging, it is also possible to estimate the temporal variation of the magnetic fields at the shock front as the CME propagates through the corona, which will allow one to understand the shock physics and its propagation in more detail.

\subsection{Spectropolarimetric Imaging of Radio CMEs}
\begin{figure}
    \centering
    \includegraphics[trim={0cm 0cm 0cm 0cm},clip,scale=0.3]{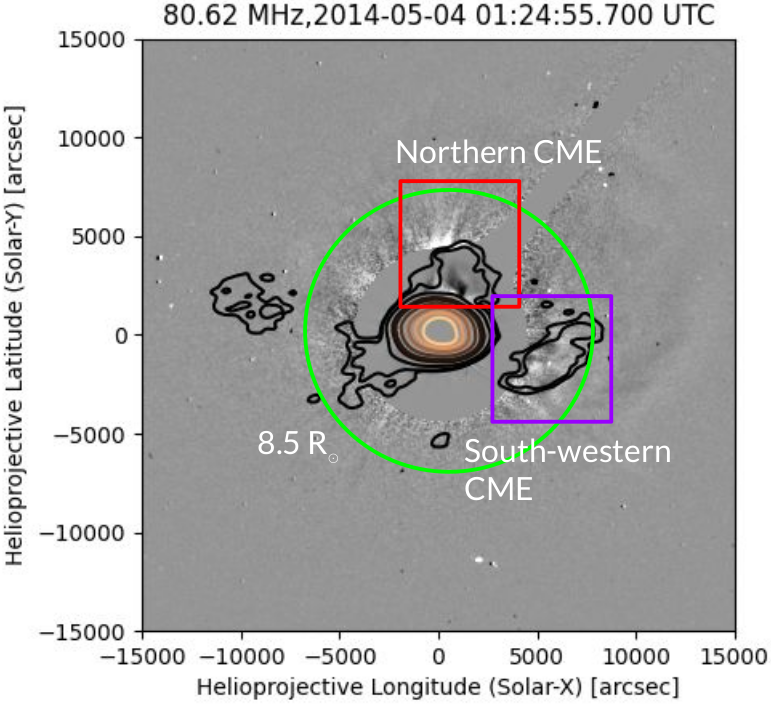}
    \vspace{-0.3cm}
    \caption{GS radio emission from two slow CMEs. CME-1 is propagating towards the North and is marked by the red box. CME-2 is propagating towards the south-western direction and is marked by the purple box.}
    \label{fig:GS_CME}
    \vspace{-0.5cm}
\end{figure}

GS emission from the CME plasma is a unique tool to measure the CME magnetic fields and other plasma parameters remotely. However, after the first detection by \cite{bastian2001} only a handful of studies \citep{Maia2007,Tun2013,Bain2014,Carley2017,Mondal2020a,Chhabra_2021} could successfully detect GS emission from CMEs. Most of the detections are from fast CMEs. \cite{Mondal2020a} demonstrated that with high-dynamic range imaging with the MWA, it is possible to detect GS emission from slow CMEs as well and also at much higher coronal heights ($\sim4.73\ R_\odot$). In another event, \cite{Kansabanik2023_CME1} detected GS emission from another two slow CMEs, as shown in Figure \ref{fig:GS_CME}. The GS radio emission from the southwestern CME is detected up to 8.5 $R_\odot$, the highest heliocentric distance to date. Studies by \cite{Mondal2020a} and \cite{Kansabanik2023_CME1} successfully demonstrated that the high-dynamic-range imaging offered by the MWA data allowed us to detect much fainter GS emissions from the slow CMEs. 

Not only it is now possible to routinely detect the GS emission, the good spectral coverage and robust polarization calibration offer strong constraints on the GS model parameters. Unlike the earlier studies, imaging observations allow one to perform spatially resolved GS modeling and estimate the GS model parameters. GS model has ten free parameters, which cannot be constraint using Stokes I spectrum alone. \cite{Kansabanik2023_CME1} have demonstrated that simultaneous use of constrains from Stokes I and V observations break some degeneracies between GS model parameters, the use of multi-vantage point coronagraph observations allow one to independently estimate the geometric parameters of the CME, significantly improve the robustness of the magnetic fields and other estimated parameters.

Since the peak of the GS spectra shift to lower frequency as the CME moves out, the MWA observations are well-suited to use this method to routinely estimate the CME-entrained magnetic field over the heliocentric heights of $\sim2-10\ R_\odot$. It is also possible to use the same method to estimate the CME-entrained magnetic field at lower coronal heights using high-dynamic-range solar imaging observations at higher frequencies. This may be possible shortly with the MeerKAT \citep{meerkar2016} operating at $\sim$580-1600 MHz. With the joint observations with the MeerKAT and the MWA, and in the future with the Square Kilometre Array \citep[SKA,][]{ska_concept}, it will be possible to routinely measure the spatially resolved CME-entrained magnetic field remotely from the lower coronal heights to $\sim10\ R_\odot$.

\subsection{Heliospheric Magnetic Field Measurements}\label{subsec:FR_obs}
As mentioned in the previous section, spectro-polarimetric modeling of GS emission can only be used to measure CME-entrained magnetic field upto $\sim$10 $R_\odot$, because beyond that spectral peak goes below the ionospheric cutoff, and also the emission becomes too faint to detect. Hence, beyond 10 $R_\odot$ the only remote sensing method to measure the CME-entrained magnetic field is the FR measurements of the background linearly polarized radio source. 

The main observable in an FR observation is the rotation ($\Delta\chi$) of the plane of polarization of a linearly polarized source when the CME crosses in front of it. $\Delta\chi$ is the product of the square of the wavelength of emission ($\lambda$) and ``Rotation Measure" \citep[RM,][]{Brentjens2005}. RM is proportional to the LoS integrated product of electron density; $n_e(\mathbf{r})$ and LoS magnetic field; $B_\parallel$. Since $\Delta\chi \propto\lambda^2$, FR effects are more pronounced at longer wavelengths even for the small changes in the RM. However, most FR observations are currently undertaken at 1-2 GHz with the Very Large Array \citep[VLA,][]{VLA2009}. As the CME expands into the heliosphere, RM decreases since both $n_e$ and $\mathbf{B}$ decrease. VLA observations can asses CME RM contributions  ($\geq1\ \mathrm{rad/m^{2}}$) up to $\sim15\mathrm{R_\odot}$ \citep{kooi2017,Kooi2021}. On the other hand, low-frequency ($\sim100$ MHz) observations can measure the RM contribution from CMEs up to 80$\mathrm{R_\odot}$ ($\sim0.01\ \mathrm{rad/m^{2}}$) \citep{Oberoi2012}.  

To arrive at the vector magnetic fields from the measured LoS integrated magnetic field using FR observations, one has to use magnetic flux rope (MFR) models of the CME. The state-of-the-art MFR models like FRiED \citep{Isavnin_2016} and 3DCORE \citep{Möstl2018} have tens of free parameters. Hence, to constrain these free parameters one needs a large numbers of LoS measurements crossing the CME \citep{Kooi2021,wood2020}. The measured RM and $n_e$ estimates from white-light observations provide the LoS integrated magnetic fields ($B_\parallel$) for each LoS. Multi-LoS $B_\parallel$ measurements will allow us to go beyond self-similar assumptions and constrain MFR models for CME deformations and thus improve forecasting of their 1 AU properties.

\begin{figure}
    \centering
    \includegraphics[trim={4cm 0cm 0cm 0cm},clip,scale=0.35]{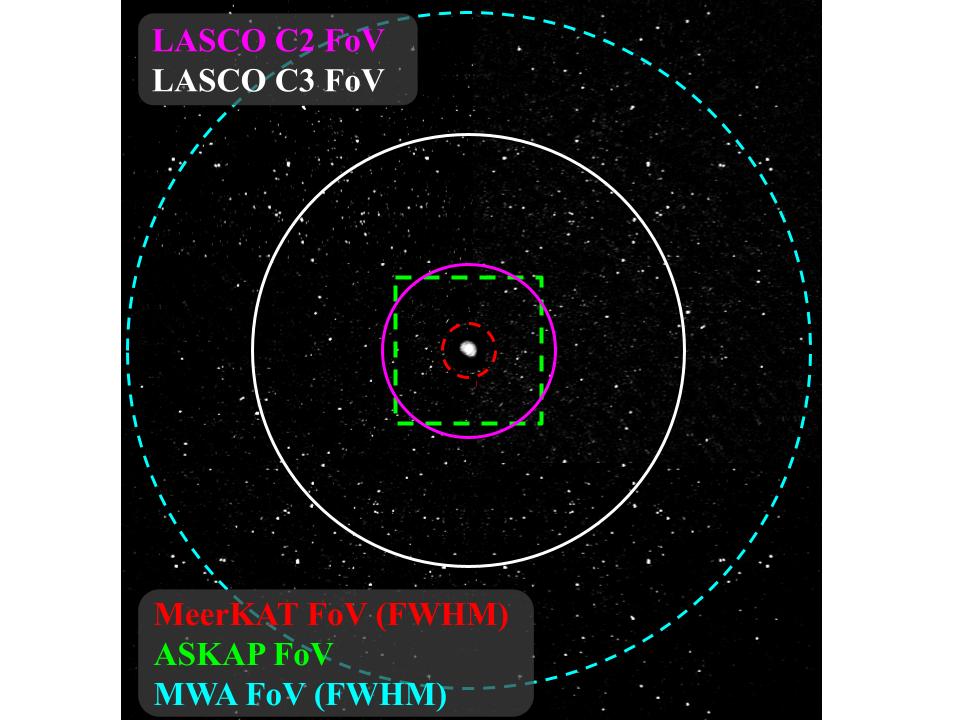}
    \caption{FoV of different new-generation wide-FoV instruments including the FoV of LASCO C2 and C3 coronagraphs are shown on top of the MWA Stokes I image. More than 80 background radio sources are visible simultaneously with the Sun in the FoV.}
    \label{fig:FR_FOV}
    \vspace{-0.5cm}
\end{figure}

Due to the small field-of-view (FoV) ($\sim$2.5$\mathrm{R_\odot}$ at 1 GHz), the earlier studies using the VLA are restricted to one background source at a time and have to cycle over a number of them multiple times during the observing campaign (4-6 hours). To derive the vector magnetic field, the analysis has to assume that the reconstructed MFR from white-light observations remains constant, and hence the observed RM time series are due to the passage of this unchanging MFR across the LoS. However, CMEs evolve with time and this evolution must be considered to improve the accuracy of CME magnetic fields measurements. The underlying assumption of constancy of MFR is the key limitation of the current state-of-the-art FR experiments. One can overcome this limitation by using spatial constraints from simultaneous FR-measurements along multiple LoS using the new-generation wide-FoV instruments like the MWA (FoV is $\sim80\ R_\odot$), MeerKAT (FoV is $\sim9\ R_\odot$) and Australian Square Kilometre Array Pathfinder \citep[ASKAP,][]{askap2021} (FoV is $\sim30\ R_\odot$). FoV of these instruments compared to LASCO \citep{Brueckner1995} coronagraphs are shown in Figure \ref{fig:FR_FOV}.

On the other hand, wide-FoV instruments come with their own challenges. Since the Sun is the strongest radio source in the sky, it can contribute to the observed background even when it lies in the sidelobes of the primary beams of the telescopes. Hence, one needs to detect the background galactic/extra-galactic sources in the presence of solar emission. Although wide-field polarimetric observations for astronomical sources are well-established for both MWA \citep{Risley2020} and MeerKAT \citep{Anderson2021}, one needs to tackle other challenges when the Sun is inside the FoV. These challenges have now been overcome using the P-AIRCARS. As a first achievement, it is now possible with the MWA to detect Stokes I emission from background galactic/extra-galactic radio sources in the presence of the Sun in the FoV \citep{Kansabanik_2021} over a small spectro-temporal integration (2 MHz and 2 minutes). In Figure \ref{fig:FR_FOV}, small white dots are background radio sources that are visible simultaneously with the Sun at the center. Another bright linearly polarized source is the galactic diffuse emission, which can also be detected simultaneously with the Sun the FoV \citep{Oberoi2022}.

With the next phase of the MWA (phase-III, with 256 antenna tiles), we hope to detect the linearly polarized emissions from the background sources to measure the heliospheric FR. We also anticipate that in near future this will also be possible with the MeerKAT and ASKAP, which provide higher (2-5 $\mathrm{deg^{-2}}$) \citep{McConnell2020} source density than the MWA ($\sim0.05\ \mathrm{deg^{-2}}$) \citep{Risley2020}, suitable for FR observations at outer coronal heights. 

\section{Benefits from Combined Radio Observations with Aditya-L1}\label{label:aditya}
In Section \ref{sec:radio_results} we have demonstrated that recent developments in radio observations now allow us to measure the magnetic field both at the CME shock front as well as inside the CME plasma at coronal heights. But, these observations also need complementary white-light observations. While LASCO C2 and C3 observations can provide observations above 2 $R_\odot$, CME observations at lower coronal heights are rather limited. 

These missing observations will be filled by the Visible Emission Line Coronagraph \citep[VELC,][]{aditya_velc} onboard the upcoming Aditya-L1 mission \citep{Tripathi2017}. The two major advantages of the VELC are its FoV and spectro-polarimetric observing capability. VELC is designed to image the solar corona from 1.05 $R_\odot$ to 3 $R_\odot$. VELC consists of a continuum channel to produce the coronagraph images in the continuum with a central wavelength close to 530 nm covering the entire FoV \citep{Nagaraju:21}. VELC also has three spectroscopy channels. The spectrograph is designed to observe the corona in three spectral lines at 530.3 nm due to Fe XIV (green channel), 789.2 nm due to Fe XI (red channel), and 1074.7 nm due to Fe XIII (IR channel). VELC spectrograph is comprised of four equispaced multi-slit spectrographs, but covering only up to 1.5 $R_\odot$. The IR channel also has a polarimeter for full Stokes spectropolarimetric observations.

The continuum observations with VELC at lower coronal heights will provide crucial information about the CME velocity and acceleration at these heights, which are important to understand the shock properties of the CMEs at these heights. Since the ground-based instruments can only offer observations 
 of type-II radio bursts up to $\sim2\ R_\odot$, VELC observations are important to have complementary information of CME dynamics at lower coronal heights. On the other hand, VELC will be the first coronagraph with a multi-slit spectro-polarimeter. In the presence of magnetic fields, the Zeeman splitting of the spectral line gives three spectral components, which are right circularly, left circularly, and linearly polarized, respectively. The unique capability of spectro-polarimetric observations will hence provide an independent measurement of both the magnetic field strength and also the direction in the plane of the sky \citep{aditya_mag_1,aditya_mag_2}. On the other hand, GS modeling provides LoS-integrated magnetic fields. Hence, combining the VELC observations, GS modeling with MeerKAT observation, and CME models can provide complete 3D magnetic field measurements of the CME plasma at the lower coronal heights. At middle coronal heights, GS emission will remain the only possible method to remotely measure the magnetic fields of the CME. 

Aditya-L1 does not only have remote sensing instruments, but it will also carry some {\it in-situ} instruments \citep{Janardhan2017,Chakrabarty2022}, namely the Aditya Solar Wind Particle Experiment (ASPEX) \citep[ASPEX,][]{Goyal2022}, the Plasma Analyser Package for Aditya \citep[PAPA,][]{Thampi2014}, and Magnetometer (MAG). While ASPEX and PAPA are particle detectors, the MAG instrument will provide {\it in-situ} magnetic field measurements from the L1 point. These {\it in-situ} measurements are essential to validate the predictions of $B_\mathrm{z}$ component of the CME magnetic field using heliospheric FR observations (Section \ref{subsec:FR_obs}), which will allow one to quantify the advantages and limitations of different CME models. 

\section{Synergies with PUNCH Mission}\label{label:punch}
Polarimeter to UNify the Corona and Heliosphere \citep[PUNCH,][]{punch2022} is a NASA Small Explorer mission to observe the corona and heliosphere from 6-180 $R_\odot$ using simultaneous observations from four spacecraft in a Sun-synchronous orbit. The main advantage of the PUNCH will be provided by its highly sensitive coronagraph, the Narrow Field Imager \citep[NFI,][]{NFI_2019} and heliospheric imagers, Wide Field Imager \citep[WFI,][]{WFI_2019} and the polarization observing capabilities of all of these instruments. 

The heliospheric FR measurements will provide the RM along each LoS. RM is the LoS integrated product of $n_e(\mathbf{r})$ and $B_\parallel$. Hence, to estimate the $B_\parallel$ from each LoS, one needs an independent measurement of $n_e$. The sensitivity of LASCO C3 only allows us to estimate $n_e$ up to $\sim15-20\ R_\odot$. On the other hand, at these heights, the F-corona starts to dominate the K-corona, which makes the $n_e$ estimation from white-light observations using the method developed by \cite{Hayes2001} inaccurate. The highly sensitive polarimetric observing capability of the PUNCH can overcome these limitations. PUNCH observations 
will allow us to measure 3D $n_e$ up to $180\ R_\odot$ \citep{DeForest2017}. PUNCH observations will also provide additional constraints to CME MFR models along with FR measurements, which will improve the accuracy of the magnetic field prediction of the CME using joint heliospheric observations using the PUNCH and other ground-based radio interferometers. 

\section{Conclusion}\label{sec:conclusion}
Over the past several years, ground-based radio observations of the Sun and heliosphere have been revolutionized. These developments in radio wavelengths have not been stopped but are continuously improving. These successes also have demonstrated the potential of the upcoming world's largest radio telescope, the SKA, for solar and heliospheric observations. 

These achievements and success at radio wavebands will be supported by some upcoming unique space-based observing facilities, like the Aditya-L1 and PUNCH. Both of these missions are going to be launched in near future. Both of these instruments have some unique and unprecedented capabilities. These observing capabilities will provide several pieces of complementary information, which will enable more fruitful use the radio observations. We anticipate all of these new-generation heliospheric observatories from visible to radio wavelengths will improve our understanding of the space weather, and enable a more accurate prediction of space weather to save our modern days technologies from space-weather hazards.

\section*{Acknowledgments}
{This scientific work makes use of the Murchison Radio-astronomy Observatory (MRO), operated by the Commonwealth Scientific and Industrial Research Organisation (CSIRO). We acknowledge the Wajarri Yamatji people as the traditional owners of the Observatory site.  Support for the operation of the MWA is provided by the Australian Government's National Collaborative Research Infrastructure Strategy (NCRIS), under a contract to Curtin University administered by Astronomy Australia Limited. We acknowledge the Pawsey Supercomputing Centre, which is supported by the Western Australian and Australian Governments. D.K., D.O. and P.M. acknowledge support of the Department of Atomic Energy, Government of India, under the project no. 12-R\&D-TFR-5.02-0700. S.M. acknowledges the
partial support from USA NSF grant AGS-1654382 and USA NASA grant 80NSSC20K1283 to the New Jersey Institute of Technology. This research has also made use of NASA's Astrophysics Data System (ADS).}

\bibliographystyle{cs20proc}
\bibliography{example.bbl}

\end{document}